\documentstyle[12pt,fleqn]{article}
\def\be{\begin{equation}}
\def\ee{\end{equation}}
\def\la{\mathrel{\mathpalette\fun <}}

\def\vev#1{{\langle #1 \rangle}}
\def\fun#1#2{\lower3.6pt\vbox{\baselineskip0pt\lineskip.9pt
        \ialign{$\mathsurround=0pt#1\hfill##\hfil$\crcr#2\crcr\sim\crcr}}}
\def\refto#1{{[\ref{#1}]}}
\def\mpl{{M_{Pl}}}
\def\prd#1#2#3{{{\it Phys.~Rev.~D}~{\bf #1} (#3) #2}}
\def\prl#1#2#3{{{\it Phys.~Rev.~Lett.}~{\bf #1} (#3) #2}}
\def\plb#1#2#3{{{\it Phys.~Lett.}~{\bf #1B} (#3) #2}}
\def\npb#1#2#3{{{\it Nucl.~Phys}~{\bf B#1} (#3) #2}}
\def\prp#1#2#3{{{\it Phys~Rep.}~{\bf #1} (#3) #2}}
\def\thetabar{\overline{\theta}}
\begin{document}
\begin{titlepage}
\null\vspace{-62pt}
\begin{flushright}NSF--ITP--92--06\\
                  CMU--HEP92--05\\
                  FNAL--PUB--92/34--A\\
                  HUTP--92A011\\
                  VAND--TH--92--2\\
                  January, 1992
\end{flushright}
\vspace{0.2in}
\centerline{{\large \bf Solutions to the Strong-CP Problem in a World with
Gravity}}
\vspace{0.3in}
\centerline{Richard Holman,$^{(a,b)}$\ \  Stephen D.~H.~Hsu,$^{(a,c)}$\ \
Thomas W.~Kephart,$^{(d)}$}
\vspace{0.2in}
\centerline{Edward W.~Kolb,$^{(a,e)}$\ \  Richard Watkins,$^{(e)}$\ \
Lawrence M.~Widrow$^{(a,f)}$}
\vspace{0.3in}
\centerline{{\it $^{(a)}$Institute for Theoretical Physics, University of
California, Santa Barbara, CA~~93106}}
\vspace{.1in}
\centerline{{\it $^{(b)}$Physics Department, Carnegie Mellon University,
Pittsburgh, PA~~ 15213}}
\vspace{.1in}
\centerline{{\it $^{(c)}$Lyman Laboratory of Physics, Harvard University,
Cambridge, MA~~ 02138}}
\vspace{.1in}
\centerline{{\it $^{(d)}$Department of Physics and Astronomy,
 Vanderbilt University, Nashville, TN ~37235}}
\vspace{.1in}
\centerline{{\it $^{(e)}$NASA/Fermilab Astrophysics Center}}
\centerline{{\it Fermi National Accelerator Laboratory, Batavia, IL~~60510}}
\centerline{{\it and Department of Astronomy and Astrophysics, Enrico Fermi
        Institute}}
\centerline{{\it The University of Chicago, Chicago, IL~~ 60637}}
\vspace{.1in}
\centerline{{\it $^{(f)}$Canadian Institute for Theoretical Astrophysics}}
\centerline{{\it University of Toronto, Toronto, Ontario, M5S 1A1, CANADA}}
\vspace{.4in}
\baselineskip=18pt
\centerline{Abstract}
\begin{quotation}
We examine various solutions of the strong-CP problem to determine their
sensitivity to possible violations of global symmetries by Planck scale
physics. While some solutions remain viable even in the face of such effects,
violations of the Peccei--Quinn (PQ) symmetry by non-renormalizable operators
of dimension less than $10$ will generally shift the value of $\thetabar$ to
values inconsistent with the experimental bound $\thetabar \la 10^{-9}$. We
show
that it is possible to construct axion models where {\em gauge} symmetries
protect PQ symmetry to the requisite level.
\end{quotation}
\end{titlepage}
\newpage

\baselineskip=24pt

\vspace{24pt}

It is well known that there are two contributions to CP violation in the
standard model.  First, QCD instantons induce a term ${\cal L}_{\rm QCD}=
\theta\, {\rm tr} \, G\widetilde{G}$ in the effective Lagrangian, which
violates both P and CP \refto{GGDUAL}. Here, $\theta$ is a dimensionless
coupling constant, which one might naively expect to be of order unity.
Second, the quark mass matrix can be complex, leading to a CP-violating phase
in the Kobayashi-Maskawa mixing matrix. The phase of the quark mass matrix
gives rise to an additional contribution $\theta_{\rm QFD} = \rm{arg} \det{\cal
M}_q$ to the coeffiecient of ${\rm tr} \, G\widetilde{G}$. The degree of
strong-CP violation is controlled by the parameter $\thetabar=\theta+\rm{arg}
\det{\cal M}_q$, which is constrained by measurements of the electric dipole
moment of the neutron to be less than $10^{-9}$ \refto{EXPLIMIT}. The strong-CP
problem is that there is no reason for these two contributions, which arise
from entirely different sectors of the standard model, to sum to zero to such
high accuracy.

The solutions that have been proposed for the strong-CP problem fall into three
general classes.  First, there are those that rely on the existence of an extra
global $U(1)_A$ symmetry. This symmetry arises naturally if one or more of the
quark masses are zero \refto{MUZERO}. In this case, it can be shown that the
QCD  $\theta$ parameter becomes unobservable. This solution is considered
unattractive, since experimental evidence implies that it is unlikely that any
of the quarks are massless. Peccei and Quinn \refto{PQ} (PQ) proposed a
solution to the strong-CP problem in which they introduced an auxiliary, chiral
$U(1)_{\rm PQ}$ symmetry that is spontaneously broken at a scale
$f_a$, giving rise to a Nambu--Goldstone boson $a$ known as the axion
\refto{WW}. This symmetry is explicitly broken by instanton effects. This
explicit breaking generates a mass for the axion of order
$m_a\sim\Lambda^2/f_a$, where $\Lambda$ is the QCD scale. The important point
is that the effective potential for the axion has its minimum at $\vev{a/f_a}
=-\thetabar$. It follows that when the axion field relaxes to its minimum, the
coefficient of ${\rm tr} \, G\widetilde{G}$ is driven to zero. This solution
has
received the most attention and has been explored by many authors.

A second class of solutions involve models where an otherwise exact CP is
either softly or spontaneously broken.  Specific models have been proposed
where $\theta$ is calculably small and within the experimental limits
\refto{SOFT}.

A third class of solutions involve the action of wormholes \refto{WORMCP}.
As we will argue below, wormholes can break global symmetries explicitly,
thus giving rise to potentially large contributions to $\thetabar$. However,
under certain assumptions, it can be shown that wormholes actually have the
effect of setting $\thetabar = 0$ \refto{WORMCP}.

In this letter, we address the question of whether these solutions to the
strong-CP problem can remain viable if Planck scale effects break global
symmetries explicitly. There are many arguments suggesting that all global
symmetries are violated at some level by gravity. First, no-hair theorems tell
us that black holes are able to swallow global charge. This allows for a
gedanken experiment in which a quanta with global charge ``scatters'' with a
black hole, leaving only a slightly more massive black hole, but one with
indeterminate global charge as dictated by the no-hair theorem. Heuristically,
if one considers ``virtual'' black hole states of mass $M$ arising from quantum
gravity, one can integrate them out to yield global charge violating operators
suppressed by powers of $M$, where $M$ might be as small as $\mpl$, the Planck
mass.

Another indication that gravity might not respect global symmetries comes from
wormhole physics \refto{WORM}. Wormholes are classical solutions to Euclidean
gravity that describe changes in topology. Integrating over all wormholes (with
a cutoff on their size) yields a low-energy effective action that contains
operators of {\em all} dimensions that violate global symmetries \refto{SJR}.
The natural scale of violation in this case is the wormhole scale, usually
thought to be very near (within an order of magnitude or so) $\mpl$.

Without explicit calculations of these effects, we are left with the following
prescription:  Due to our lack of understanding of physics at the Planck scale,
we have no choice but to interpret theories that do not include gravity in a
quantum mechanically consistent way as {\em effective} field theories with a
cutoff at $\mpl$. If we adhere rigorously to this principle, we are then
required to add all higher dimension operators (suppressed by powers of
$\mpl$) consistent with the symmetries of the full theory at $\mpl$.  As
discussed above, it seems very unlikely that the full theory respects global
symmetries.  We note that it would be particularly surprising if the entire
theory respects $U(1)_{\rm PQ}$, since this symmetry is already explicitly
broken by instanton effects. We should note that similar ideas were noted
briefly in the prescient paper of Georgi, Hall, and Wise \refto{GEORGIETAL};
however, we are now in a position to be somewhat more specific about the nature
of the Planck scale effects in question and to explore their consequences.

We consider first the implications for the axion model. To be specific, we
consider a generic invisible axion model \refto{DFS} in which an electroweak
singlet $\phi$, charged under $U(1)_{\rm PQ}$, is responsible for spontaneous
breaking of the PQ symmetry. We may parametrize $\phi$ on the vacuum manifold
as $\phi = (f_a/\sqrt{2})\ \exp(ia/f_a)$, where $a$ is the axion field.
The effects of the QCD anomaly are to generate a mass
for the axion of order $m_a \sim \Lambda^2 / f_a$, where $\Lambda$ is the QCD
scale. A variety of astrophysical
and cosmological constraints on the axion force $f_a$ into a narrow range of
$10^9{\rm GeV}\la f_a \la 10^{12}{\rm GeV}$ for standard axions,
or in a still narrower range around $10^7$GeV for hadronic axions \refto{KT}.

The instanton induced potential for $a$ takes the form \refto{PQ}:
\be
\label{eq:VPQ}
V(a) = \Lambda^4 \cos( a/f_a + \theta).
\ee
where $\theta$ is the QCD
theta angle in a basis where the quark mass matrix is real.
While dominating the path integral with instantons is probably a bad
approximation in an unbroken gauge theory like QCD, there are rigorous results
\refto{VAFAWITTEN} showing that the minimum of $V(a)$ occurs at strong-CP
conserving values.

One possibility is that gravity does not respect $U(1)_{\rm PQ}$ at all, as is
the case if wormhole effects are large. In this case, one should include
renormalizable operators such as
\be
\Delta V(\phi) \sim M_W^2\phi^2 + \rm{h.c.}
\ee
Here $M_W$ is the
wormhole scale, which is expected to be of the order of the Planck mass. With
the addition of these operators, the PQ symmetry is strongly broken and axions
never arise at all.

A second possibility is the $U(1)_{\rm PQ}$ is only broken through
non-renormalizable operators of higher dimension.  This can occur if either
wormhole effects are suppressed or if the PQ symmetry is automatic, i.e.,
it is present ``automatically'' when one includes all renormalizable terms
consistent with a given gauge group.  As we shall see below, higher dimension
operators will also spoil the axion solution to the strong-CP problem except
possibly in
the case of an automatic PQ symmetry, where gauge symmetries can eliminate
operators up to some required high dimension.

We now explore the effect upon the axion potential of dimension $D$ operators
such as
\be
{\cal O}_{D} = \frac{\alpha_{D}}{\mpl^{D-4}} ~
                {\phi^{*}}^a \phi^b+{\rm h.c.} \qquad \qquad (a\ne b;~~a+b=D),
\ee
which explicitly break $U(1)_{PQ}$. Operators of dimension $D$ will modify the
axion potential of Eq.~(\ref{eq:VPQ}):
\be
V(a) =  \Lambda^4 \cos( a/f_a + \theta) + \sum {\Delta_n \cos(n
        a/f_a + \delta_n) } \qquad (n=D,~D-2,~D-4,\ldots),
\ee
where $\Delta_n \sim \alpha_D f_a^{D}/\mpl^{D-4}$, and $\delta_n$ is a phase
angle.  Let us simply analyze the $n=1$ contribution.  The extra contribution
will shift the minimum of the axion potential away from the strong-CP
conserving minimum of $\vev{a/f_a}=-\theta$.  Unless
$\epsilon = \vev{a/f_a}+\theta$ is less than $10^{-9}$ the amount of CP
violation obtained will be in conflict with experiment. The minimum of the
axion potential is now determined by $f_aV'(a)\simeq\Lambda^4\epsilon +\Delta_1
\sin(\epsilon-\theta+\delta_1)=0$. The magnitude of $\sin(\epsilon-\theta +
\delta_1)$ will not, in general, be small, and $\epsilon \sim
\Delta_1/\Lambda^4$.

Since we know $\epsilon < 10^{-9}$, $\Delta_1<10^{-9}\Lambda^4$.  For dimension
$D$ operators, we expect $\Delta_1\sim \alpha_D f_a^{D}/\mpl^{D-4}$. Using
$\Lambda=10^{-1}$GeV, the limit on $\epsilon$ translates into the following
limit on the dimension $D$ of the operator as a function of $f_a$ and
$\alpha_D$:
\be
\label{eq:DFA}
D \la \frac{89+\log\alpha_D}
        {9-\log{(f_a/10^{10}{\rm GeV})}}.
\ee
If Eq.~(\ref{eq:DFA}) is satisfied, it is very simple to show that the
higher-dimension operators will have an insignificant effect on the axion mass.
In fact, the zero temperature axion mass is just $m_a \sim \Lambda^2 (1 +
\epsilon)/f$. However, we should note that the temperature dependence of the
axion mass is quite different in the presence of higher dimensional operators.
In particular, the mass induced by the higher dimension operators is {\em
always} ``turned on.'' This may affect axion cosmology in interesting ways. We
are currently investigating this topic, as well as such effects on other
theories (such as Majoron models) relying upon Nambu--Goldstone boson physics
\refto{US}.

These results at first seem puzzling, since low-energy physics is not in
general sensitive to physics at the Planck scale.  However, Nambu--Goldstone
bosons have the peculiar property that although they are massless (or very
light in the case of pseudo-Nambu--Goldstone bosons such as the axion), they
are
not, properly speaking, part of the low-energy theory as evidenced by the fact
that self-couplings, and couplings to light fields are suppressed by a power of
a large mass scale.  The fact that a light particle such as the axion is part
of the high-energy sector accounts for its interesting properties, but also
renders it susceptible to high-energy corrections.

In a generic invisible-axion model, there is no reason why a term such as
$\phi^5/\mpl$ could not be generated (here $\phi$ is a gauge-singlet field).
This term would give rise to unacceptable shifts in $\thetabar$ unless
$\alpha_D \la 10^{-44-\log(f_a/10^{10}{\rm GeV})}$, which is remarkably small.
Is there any to avoid this problem?

There are, in fact, ways to construct axion models which suppress higher
dimensional operators as needed. This construction is based on the notion of
automatic PQ symmetries \refto{GEORGIETAL}, as described above. We first
consider a supersymmetric automatic model based on the gauge group $E_6 \times
U(1)_X$ \refto{ESIX}. The superfield content of the model is some number of
$\bf 27$'s with $X$ charges $\pm 1$ and a $\overline{\bf 351}$ with $X$ charge
$0$. The most general renormalizable, gauge-invariant superpotential will only
contain terms of the form ${\bf 27}_1\cdot{\bf 27}_{-1}\cdot\overline{\bf
351}_0$, where the subscripts denote the $U(1)_X$ charges. This automatically
gives rise to a PQ symmetry in which the $\bf 27$'s have PQ charge $+1$ and the
$\overline{\bf 351}$ has PQ charge $-2$. The lowest dimension operators
consistent with gauge invariance in the superpotential that break the PQ
symmetry are terms like ${\bf 27}^{\,6}$, $\overline{\bf 351}^{\,6}$, and
$({\bf 27}\cdot{\bf 27}\cdot\overline{\bf 351})^4$. These will then give rise
to dimension $10$ operators in the effective Lagrangian. Furthermore, it is
relatively easy to see that we can break the gauge symmetries and the PQ
symmetry spontaneously in such a way so that the final PQ symmetry (a linear
combination of the original PQ symmetry and some broken gauge symmetries) is
broken around $10^{10}$ GeV.

It is also possible to construct automatic PQ models based on supersymmetric
$SU(N)$ GUT's that suppress higher dimension operators to any desired level for
sufficiently large $N$. Models of this type without exotic fermions must all
have at least four different chiral matter irreducible representations whose
Young tableaux consist of a single column. Needless to say, these are
exceedingly unattractive models. They will tend to have many extra families,
which in addition to a host of phenomenological problems, will possibly destroy
the asymptotic freedom of QCD.

Planck scale physics may also significantly affect the other solutions for the
strong-CP problem \refto{JEKIM}.  As described above, the second class of
solutions are based upon models where CP is softly or spontaneously broken. How
they fare under Planck scale physics depends on whether dimension four
operators are generated, or whether only higher dimension operators appear. If
renormalizable operators can be generated, then the violation of CP by Planck
scale effects will give rise to a ${\rm tr} \, G\widetilde{G}$ term, thus
regenerating the strong-CP problem (we should note, however, that the
coefficient of such a term could be exponentially suppressed if it appeared in
some controlled semiclassical expansion about some classical configuration
\refto{SJR}).

Let us next consider the case in which only non-renormalizable operators are
generated by Planckian physics. In this case, all models with fields that
acquire vacuum expectation values well below the Planck scale (typically the
weak scale), will generate corrections to $\thetabar$ that are highly
suppressed by powers of $\mpl$. In essence, this is nothing more than a
restatement of the effective field theory philosophy: as long as we consider
physics at energies below the cutoff of our theory, the dominant effects come
from the renormalizable operators in the theory. This way of thinking about
effective field theories explains why the PQ solution is so susceptible to
possible effects of gravity. The problem is that the PQ scale is too close to
$\mpl$ while the constraints on $\thetabar$ are too tight.

Although we have seen that wormholes are troublesome for models that claim to
solve the strong-CP problem, there is some indication that wormhole effects
themselves might drive the QCD $\thetabar$ parameter to a CP conserving value
\refto{WORMCP}. Within the framework of Coleman's wormhole calculus
\refto{COLWORM} (which has since been shown to be naive in some respects
\refto{SDK}), $\thetabar$ became a function of the wormhole parameters. The
implementation of Coleman's prescription for determining the value of these
parameters was then shown to set $\thetabar$ to a CP conserving value. It is
not impossible that a more sophisticated approach to the wormhole calculus
would still lead to a similar situation. However, until a better understanding
of wormholes and quantum gravity in general is reached, this will remain a
conjecture.

In conclusion, we see that Planck-scale physics can have dramatic effects on
axion physics. If one wants to pursue the axion solution to the strong-CP
problem, automatic models such as those presented here are probably the only
consistent approach that can be taken. We have also argued that the other known
solutions are essentially unaffected by gravity. The essential difference
between the PQ and the non-axionic solutions is due to the sensitivity of the
Nambu--Goldstone boson to physics at energies near the scale of spontaneous
symmetry breaking. It remains to be seen whether other facets of the axion
scenario, such as the axion energy density crisis \refto{CRISIS} will be
modified by the effects considered here.

In the course of this work we learned that the effect of gravity on the
Peccei--Quinn mechanism is also being considered by Kamionkowski and
March-Russell \refto{SHS}, and by Barr and Seckel \refto{BARSEK}. We would like
to thank them for calling their work to our attention.

It is a pleasure to thank S.~Giddings, S.~J.~Rey and  A.~Strominger for useful
discussions. This research was supported in part by the National Science
Foundation under grant No. PHY89--04035. SDH acknowledges support from the
National Science Foundation under grant NSF--PHY--87--14654, the state of Texas
under grant TNRLC--RGFY106, and from the Harvard Society of Fellows. EWK and RW
were supported by the NASA (through grant NAGW--2381 at Fermilab) and by the
DOE (at Chicago and Fermilab), RH was supported in part by DOE grant
DE--AC02--76ER3066. TWK was supported by the DOE (grant DE--FG05--85ER40226).

\vspace{36pt}
\centerline{\bf References}
\frenchspacing

\begin{enumerate}

\item\label{GGDUAL} C. G. Callan Jr., R. F. Dashen, and D. J. Gross,
                                \plb{63}{334}{1976};\\
                    R. Jackiw and C. Rebbi, \prl{37}{172}{1976};\\
                    G. 't Hooft \prd{14}{3432}{1976}.
\item\label{EXPLIMIT} K. F. Smith, et. al., \npb{234}{191}{1990}.
\item\label{MUZERO} See, e.g., D. B. Kaplan and A. Manohar,
                                \prl{56}{2004}{1986}.
\item\label{PQ} R. D. Peccei and H. R. Quinn, \prl{38}{1440}{1977};
                                \prd{16}{1791}{1977}.
\item\label{WW} S. Weinberg, \prl{40}{223}{1978};\\
                F. Wilczek, \prl{40}{279}{1978}.
\item\label{SOFT} A. Nelson, \plb{136}{387}{1984};\\
                  S. M. Barr, \prl{53}{329}{1984}.
\item\label{WORMCP} H. B. Nielsen and M. Ninomiya, \prl{62}{1429}{1989};\\
                    K. Choi and R. Holman, \prl{62}{2575}{1989};\\
                    J. Preskill, S. P. Trivedi, and M. Wise,
                                \plb{223}{26}{1989}.
\item\label{WORM} S. Giddings and A. Strominger, \npb{307}{854}{1988};\\
                  S. Coleman, \npb{310}{643}{1988};\\
                  G. Gilbert, \npb{328}{159}{1988}.
\item\label{SJR} S. J. Rey, \prd{39}{3185}{1989}.
\item\label{GEORGIETAL} H. Georgi, L. J. Hall, and M. B. Wise,
                                \npb{192}{409}{1981}.
\item\label{DFS} M. Dine, W. Fischler, and M. Srednicki,
\plb{109}{199}{1981};\\
                 A. P. Zhitnitskii, {\it Sov. J. Nucl. Phys.} {\bf 31} (1980)
                                260.
\item\label{KT} E. W. Kolb and M. S. Turner, {\em The Early Universe}
                                (Addison--Wesley, Redwood City, 1990).
\item\label{VAFAWITTEN} C. Vafa and E. Witten, \prl{53}{535}{1984}.
\item\label{US} R. Holman, S. D. H. Hsu, E. W. Kolb, R. Watkins, and L. M.
                                Widrow, work in progress.
\item\label{ESIX} P. H. Frampton and T. W. Kephart, \prd{25}{1459}{1982};\\
                  R. Holman and T. W. Kephart, \plb{167}{169}{1986}.
\item\label{JEKIM} see J. E. Kim, \prp{150}{1}{1987}, and references therein.
\item\label{COLWORM} E. Baum, \plb{133}{185}{1983};\\
                     S. W. Hawking, \plb{134}{403}{1984};\\
                     S. Coleman, \npb{310}{643}{1984}.
\item\label{SDK} W. Fischler, I. Klebanov, J. Polchinski, and L. Susskind,
                                \npb{327}{157}{1989}.
\item\label{CRISIS} J. Preskill, M. Wise, and F. Wilczek,
                                \plb{120}{127}{1983};\\
                    L. Abbott and P. Sikivie, {\it ibid}, (1983) 133;\\
                    M. Dine and W. Fischler, {\it ibid}, (1983) 137.
\item\label{SHS} M. Kamionkowski, J. M. Russell, IASSNS-HEP-92/6.
\item\label{BARSEK} S. Barr and D. Seckel, Bartol preprint.
\end{enumerate}
\end{document}